\documentclass[twocolumn,prl,floatfix,superscriptaddress]{revtex4-1}

\usepackage[labelfont=bf,justification=raggedright,singlelinecheck=false]{caption}
\usepackage{wrapfig, graphicx, subfig}

\usepackage[nottoc, numbib]{tocbibind}

\usepackage[utf8]{inputenc}

\usepackage{physics}

\usepackage{amssymb, amsthm}

\usepackage{mathbbol}

\usepackage{siunitx}

\usepackage{nicefrac}

\usepackage{hyperref}
\hypersetup{
    colorlinks = false,
    pdfborder = {0 0 0 [3 3]}
}

\usepackage{tikz}

\begin{document}

\title{Sieve of Eratosthenes for Bose-Einstein Condensates in Optical Moir\'e Lattices}

\author{Dmitry Kouznetsov}
\affiliation{KU Leuven, Dept. of Physics and Astronomy, Research unit Quantum Solid-State Physics, B-3001 Leuven, Belgium}
\affiliation{imec, Kapeldreef 75, B-3001 Leuven, Belgium}

\author{Pol Van Dorpe}
\affiliation{imec, Kapeldreef 75, B-3001 Leuven, Belgium}
\affiliation{KU Leuven, Dept. of Physics and Astronomy, Research unit Quantum Solid-State Physics, B-3001 Leuven, Belgium}

\author{Niels Verellen}
\affiliation{imec, Kapeldreef 75, B-3001 Leuven, Belgium}

\date{\today}

\begin{abstract}
    We catalog known optical moir\'e lattices and uncover exotic lattice configurations following a geometric analog of the ancient sieve of Eratosthenes algorithm for finding prime numbers.
Rich dynamics of Bose-Einstein condensates loaded into these optical lattices is revealed from numerical simulations of time-of-flight interference patterns.
    What sets this method apart is the ability to tune the periodicity of the optical lattices without changing the wavelength of the laser, yet maintaining the local potential at the individual lattices sites.
    In addition, we discuss the ability to spatially translate the optical lattice through applying a structured phase only.
\end{abstract}

\maketitle

\textbf{Introduction} --
Optical lattices (OLs) generated by standing wave laser fields form a potential landscape into which ultracold atoms can be loaded.
The formation of a Bose-Einstein Condensate (BEC) of bosonic atoms under low temperatures subject to the symmetries of an OL enables the study of fundamental models of many-body systems~\cite{jaksch_cold_1998, greiner_quantum_2002,lewenstein_ultracold_2007,greiner_optical_2008}, facilitates observation of coherent phenomena such as Bloch oscillations~\cite{dahan_bloch_1996} or Wannier-Stark ladders~\cite{wilkinson_observation_1996}, and serves as a possible implementation of quantum computation~\cite{mandel_controlled_2003}.

Over the years, a zoo of OLs has been accumulated due to the myriad ways in which laser beam configurations can be arranged to form the trapping potentials.
Aside from the familiar square or triangular OLs~\cite{grynberg_cold_2001}, quasiperiodic potentials recently piqued the interest of researchers due to the expected hybrid crystalline/amorphous features~\cite{spurrier_semiclassical_2018,corcovilos_two_dimensional_2019,viebahn_matter_wave_2019}.
Another example is the Kagome lattice~\cite{isakov_2006,jo_ultracold_2012} which adds geometric frustration to the system and can lead to the appearance of a flat band.
However, these lattices and the dynamics of the atomic system are typically treated on an individual basis.

The common parameter of periodic OLs is the periodicity $d$, and it is a key parameter that dictates the dynamics of the ultracold atom system.
From moir\'e theory~\cite{Amidror_2009} the definition of periodicity can be extended to quasi-periodic OLs in terms of dominant frequency components.
The periodicity is usually tuned by changing the wavelength of the laser.
However, the atom-lattice interaction depends on the detuning between the atomic transition frequency and the laser field frequency.
The effect of changing the wavelength of the laser can be as drastic as changing the behavior at the high intensity foci of the OL from trapping atoms to repelling them~\cite{bloch_many_body_2008}.
For one-dimensional systems the lattice periodicity can be modified without changing the wavelength by controlling the relative angle of the two interfering laser beams~\cite{fallani_bose_einstein_2005}, but in a two-dimensional system this approach has the adverse effect of also changing the lattice symmetry.
Moreover, these approaches stretch the individual lattice sites, modifying the local gradient of the potential.
This, in turn, changes the number of atoms that can be loaded per site and thus distorts the engineered dynamics of the system.
Therefore, a framework for designing OLs with varying periodicity, yet diffraction limited foci is very valuable.
As an additional benefit, such a framework would catalog known OLs and pave the way for designing more exotic potential landscapes.

Our recently developed integer lattice method \cite{kouznetsov_revival_2020} describes an algorithmic approach to computing the orientations of laser beams for generation of OLs with variable periodicity and symmetry.
This technique combines prime number factorization in the complex plane with moir\'e theory to compute wavevector components that can be used to generate desired interference patterns.
In this letter, we consider writing the OL potential in terms of integer factorization over a number field as defined by the integer lattice method.
Our simulations of BECs in such OLs reproduce matter-wave interaction patterns reported in literature and naturally lead to a classification of OLs according to the key parameter in the integer lattice method: the field norm.
We demonstrate the utility of the developed classification by calculating the set of OLs that would allow for monochromatic tuning of the OL periodicity.

Since the integer lattice method is strongly linked to the moir\'e effect, the proposed classification scheme shows promise in other systems with controllable superlattice ordering, such as twisted multi-layer graphene~\cite{cao_correlated_2018, phinney_strong_2021, gadelha_localization_2021}, twisted van der Waals materials~\cite{wu_hubbard_2018, wang_2020, andersen_2021} and Anderson localization in photonic moir\'e lattices~\cite{wang_localization_2020}.

\begin{figure}[!ht]
	\centering
	\includegraphics[width=\linewidth]{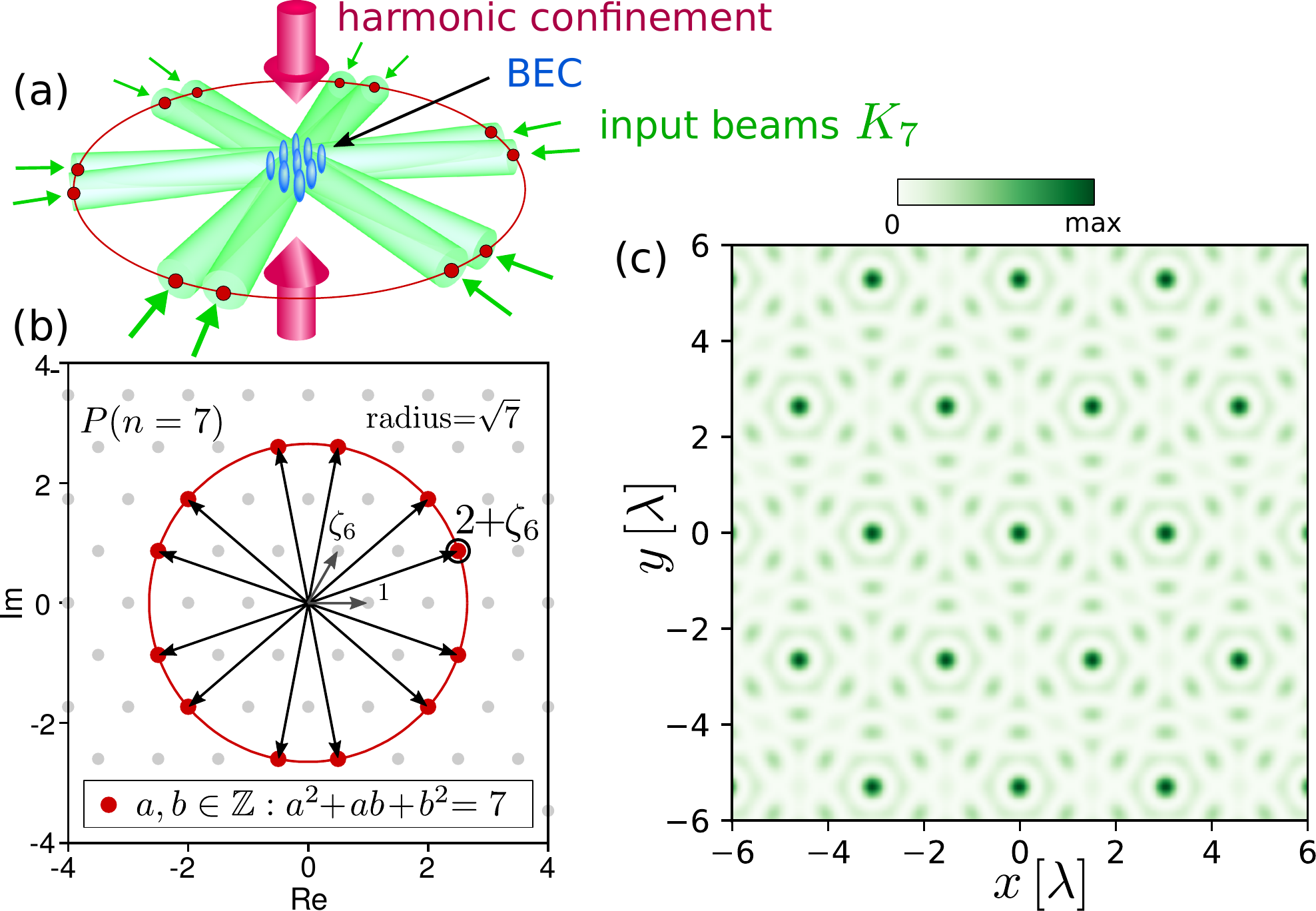}
	\caption{Generation of OLs using the integer lattice method. A planar laser beam configuration (a) is arranged such that the input beam orientations correspond to the concyclic points in the triangular integer lattice $\mathbb{Z}[\zeta_6]$ (b). The resulting interference pattern in the overlapping region of the input beams (c) shows a triangular symmetry.}
	\label{fig:mbi}
\end{figure}

\textbf{Optical lattice potential in the integer lattice method formalism} --
Before writing down the governing wavefunction of the system of cold atoms in an OL, we will integrate the integer lattice method description of coherent lattices into vector notation of the optical potential.
In Ref.~\cite{kouznetsov_revival_2020}, the set of orientations of the wavevectors of each light beam component was introduced as
\begin{equation}\label{eq:points}
	P(n)=\{N(\alpha)=n\,\mid\,\alpha\in\mathbb{Z}[\zeta_m]\}\,,
\end{equation}
where the ring $\mathbb{Z}[\zeta_m]=\{a+b\zeta_m\mid a,b\in\mathbb{Z}\}$ determines the symmetry of the system by the choice of integer $m$ in $\zeta_m=e^{2\pi i/m}$.
The corresponding field norm $N(\alpha)=\alpha\bar\alpha=n$ then selects only specific wavevectors with an integer magnitude $n$.
However, the set $P(n)$ in Eq.~\eqref{eq:points} contains only complex numbers and therefore, to give physical meaning to these elements we introduce the set
\begin{equation}\label{eq:wavevectors}
	K_n\equiv\frac{2\pi}{\lambda}\text{vec}\left(\frac{P(n)}{\sqrt{n}}\right)\,,
\end{equation}
where $\text{vec}:x+yi\mapsto(x,y)$ simply converts the complex number to vector notation.
A planar arrangement of light beams is the simplest configuration to form two-dimensional patterns and is shown in Fig.~\ref{fig:mbi}a.
The way in which the orientations of these beams is determined from the points $P(n)$ in the complex plane is illustrated in Fig.~\ref{fig:mbi}b.
Note that the points $P(n)$ are normalized to have unit length in Eq.~\eqref{eq:wavevectors} and are subsequently scaled by the wavelength $\lambda$ of the light beams.

By introducing $K_n$, we can interface the integer lattice method with the description of the general optical potential for interfering laser beams given by~\cite{popov_1987}
\begin{equation}\label{eq:potential_classic}
	V_{latt}(\vb{r})=V_0\left|\sum_j\mathcal{E}_j\boldsymbol{\epsilon}_j e^{-i(\vb{k}_j\cdot\vb{r} + \varphi_j)}\right|^2\,,
\end{equation}
where $\vb{r}$ is the position vector in two dimensions, $V_0$ is the overall strength of the potential, $\mathcal{E}_j\in[0,1]$ is the relative intensity of the laser beam, $\boldsymbol{\epsilon}_j$ is the polarization and $\varphi_j$ the phase.
The wavevectors $\vb{k}_j$ can now be replaced by the calculated wavectors in $K_n$ as follows:
\begin{equation}\label{eq:potential_new}
	V_{latt}^\prime(\vb{r})=V_0\left|\sum_{\vb{k}_j\in K_n} e^{-i(\vb{k}_j\cdot\vb{r} + \varphi_j)}\right|^2\,,
\end{equation}
assuming all beams are linearly polarized in the same direction such that $\boldsymbol{\epsilon}_j$ disappears, and have equal intensity, i.e. $\mathcal{E}_j=1$.
The phase $\varphi_j$ will come into play when discussing the translation of the OL.
For now, we set all the laser beams to be in phase, i.e. $\varphi_j=0$.
An example of a triangular OL generated from $P(7)$ is shown in Fig.~\ref{fig:mbi}c.
This final form of the optical potential facilitates the analysis of BECs loaded in OLs in the context of the integer lattice method.

\textbf{Matter-wave interference pattern simulation} --
The momentum distribution of the BEC holds key information of the system and is most often experimentally obtained by observing matter-wave diffraction in time-of-flight imaging.
We therefore target simulating matter-wave interference patterns in the following discussion.

Aside from the OL potential, we add a harmonic confinement to the BEC, such that the total potential becomes
\begin{equation}\label{eq:potential_total}
	V_{tot}(\vb{r})=V_{latt}^\prime(\vb{r})+\frac{1}{2}\omega|\vb{r}|^2\,,
\end{equation}
with trapping frequency $\omega$ in units of the recoil frequency $\omega_R=\hbar|\vb{k}|^2/2m$ with $m$ the atomic mass.
Therefore, $V_0$ in Eq.~\eqref{eq:potential_new} is in units of the recoil energy $E_R=\omega_R\hbar$.
For such a system, the weakly interacting bosonic gas is known to be well described by the time dependent Gross-Pitaevskii equation (GPE)~\cite{dalfovo_theory_1999}, which is written in dimensionless form as
\begin{equation}\label{eq:gpe}
	i\pdv{\Psi(\vb{r}, t)}{t} = \left( -\frac{1}{2}\vb{\nabla}^2 + V_{\text{tot}}(\vb{r}) + g|\Psi(\vb{r}, t)|^2 \right)\Psi(\vb{r}, t)\,,
\end{equation}
with $g$ being the variable interaction strength parameter.
We find the ground state of the system described by Eq.~\eqref{eq:gpe} using an imaginary time evolution with the Fourier split-step operator method~\cite{gaunt_2015} with absorbing boundary conditions implemented using the QuantumOptics.jl framework~\cite{kramer_quantumopticsjl_2018}.

\begin{figure*}[!ht]
	\centering
	\includegraphics[width=\linewidth]{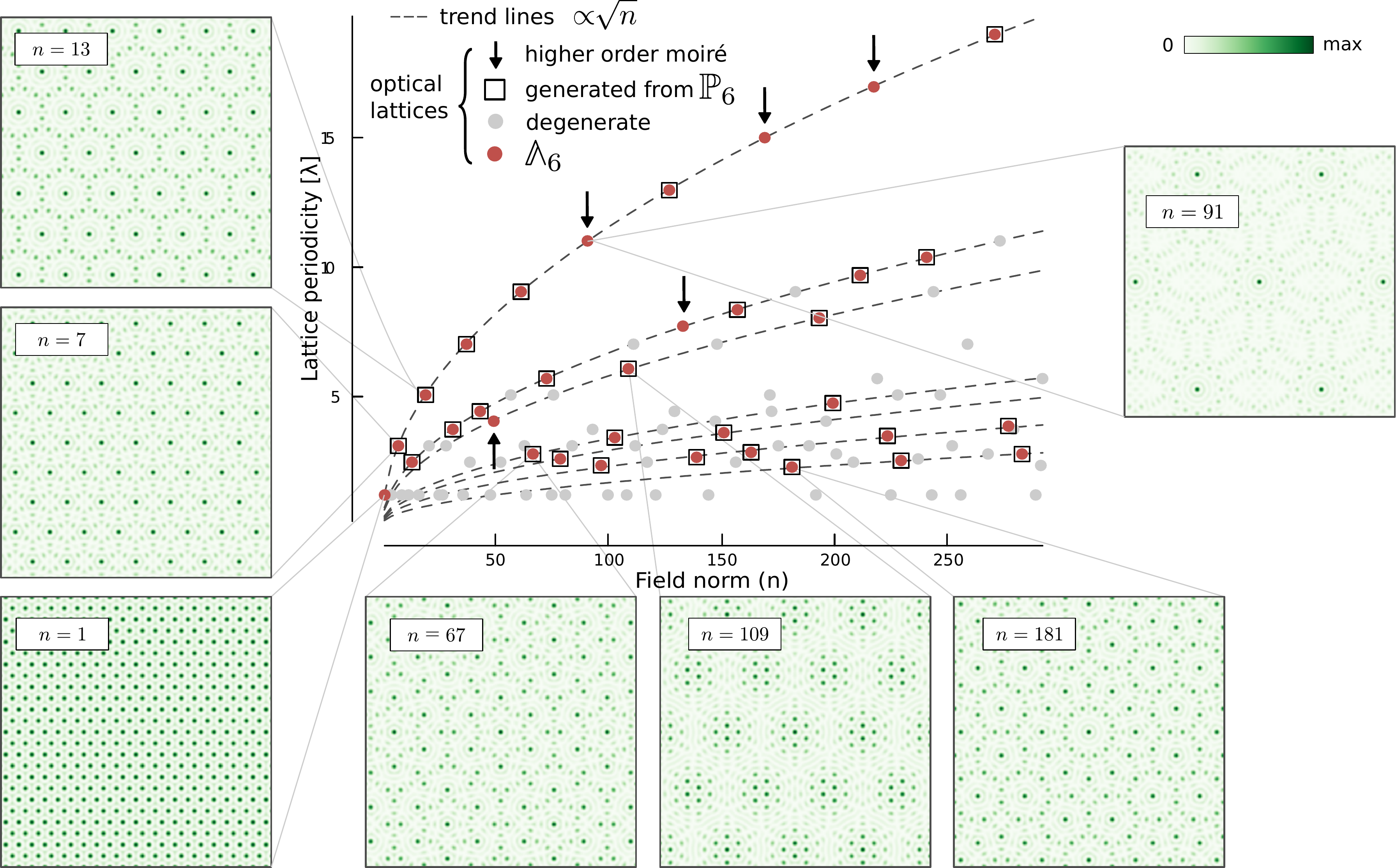}
	\caption{Integer lattice method classification of OLs analogous to the sieve of Eratosthenes algorithm. Each point in the diagram corresponds to a triangular $(m=6)$ OL generated from an integer $n$. The distinction between lattices is based on lattice periodicity, with first occuring lattices $\boldsymbol{\mathbb{\Lambda}}_6$ marked red and identical lattices marked gray. The trend lines of the periodicity are added as a visual aid (dashed lines). Non-inert prime numbers $\mathbb{P}_6$ coincide with $n$ of $\boldsymbol{\mathbb{\Lambda}}_6$ (black squares). Higher order moir\'e OLs (arrows) are generated from products of prime numbers, e.g. $91=7\cdot13$.}
	\label{fig:Eratosthenes}
\end{figure*}

\textbf{Sieve of Eratosthenes} --
Having all components for simulating bosonic gasses in OLs in place, we turn to a number theory technique to algorithmically analyze two-dimensional OL configurations.
A good place to start is the result from the integer lattice method linking coherent OLs to moir\'e superlattices with periodicity determined by the prime number factorization of an integer $n$ via Eq.~\eqref{eq:points} in the complex plane~\cite{kouznetsov_revival_2020}.
The major distinguising factor between OLs is their symmetry, which is fixed by choosing $m$.
Furthermore, from moir\'e theory it is known that the dominant spatial features are determined by the smallest components in momentum space~\cite{Amidror_2009}, such that for $\vb{k}_i\in K_n$ the pattern periodicity $d$ can be written as
\begin{equation}\label{eq:periodicity}
	d=\min|\vb{k}_i-\vb{k}_j|^{-1}\,.
\end{equation}
This quantity will act as the secondary distinguishing factor between the generated patterns.
To show that this distinction is sufficient, we consider the distribution of the set of prime numbers, which we will denote by $\mathbb{P}=\{p\in\mathbb{N}\mid p\text{ is prime}\}$.
An important observation is that not all primes remain prime in the complex plane.
For example, $5=(2+i)(2-i)$ is no longer prime in $\mathbb{Z}[\zeta_4]$.
This has important consequences for the generated coherent lattices.

Suppose a field norm $n\in\mathbb{P}$ remains prime (known as inert primes~\cite{Rotman_2013}) in $\mathbb{Z}[\zeta_m]$, then $P(n)=\emptyset$, since these numbers cannot be split in the complex plane.
Let $\mathbb{I}_m$ be the set of all such inert primes in $\mathbb{Z}[\zeta_m]$.
We can then filter out all these prime numbers that have no valid field norm assossiated with them, and be left with the set
\begin{equation}\label{eq:primes}
	\mathbb{P}_m=\mathbb{P}\backslash\mathbb{I}_m\,.
\end{equation}
The unique factorization theorem states that positive integers greater than zero can be represented in exactly one way as a product of prime numbers -- essentially describing primes as building blocks of the natural numbers.
Similarly, the factorization of $n\in\mathbb{P}_m$ in $\mathbb{Z}[\zeta_m]$ cannot be a composite decomposition of other field norms due to its primeness.
We therefore can leverage the principle of the sieve of Eratosthenes -- an algorithm in which all multiples of a number are marked iteratively such that only all primes remain~\cite{Greaves_1997} -- to identify all OLs which appear for the first time, i.e. have the lowest $n$ for a given periodicity $d$.
These lattices will be labeled $\mathbb{\Lambda}_m$.

In Fig.~\ref{fig:Eratosthenes}, each red dot corresponds to one of the first occuring lattices $\Lambda_n\in\mathbb{\Lambda}_6$.
Lattices generated from different field norm values $n$ are said to be degenerate if they have the same periodicity $d$ (see Eq.~\eqref{eq:periodicity}).
These duplicate lattices are iteratively marked grey.
For example, one can easily verify that solving Eq.~\eqref{eq:points} for $n=1$ and $n=4$ results in the same set of wavevectors due to the normalization factor  $\sqrt{n}$ in Eq.~\eqref{eq:wavevectors}, e.g. $K_1=K_4$.
Also, the trend lines of the periodicity are proportional to $\sqrt{n}$, and are plotted as a visual aid.
From construction, any lattice $\Lambda_n$ generated from $n\in\mathbb{P}_6$ will be in $\mathbb{\Lambda}_6$.
However, some lattices are constructed from a product of prime numbers (highlighted with arrows in Fig.~\ref{fig:Eratosthenes}), and will appear for the first time.
For example $\Lambda_{91}$, where $n=91=13\cdot7$.
In Ref.~\cite{kouznetsov_revival_2020} these were identified as higher order moir\'e superlattices, since their construction is a superposition of past lattices.
Of course, the exception is $\Lambda_1$, since $1$ is not a prime number.
It is important to note that even though $\Lambda_n$ have varying periodicity, the foci are diffraction limited.
\begin{figure}[!ht]
	\centering
	\includegraphics[width=\linewidth]{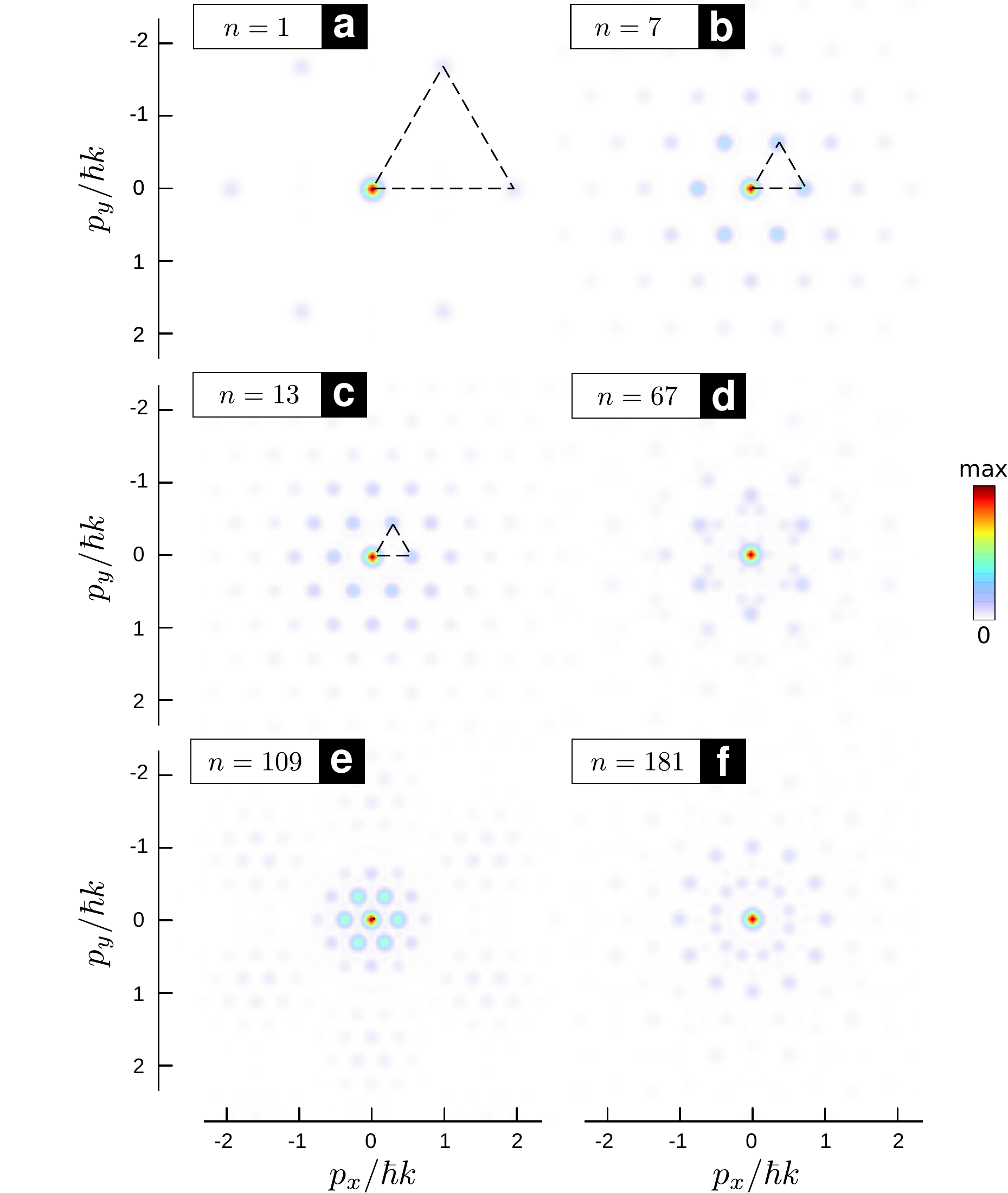}
	\caption{Calculated matter-wave interference patterns corresponding to the momentum distribution of the superfluid in the OLs obtained using the integer lattice method (see Fig.~\ref{fig:Eratosthenes}). Each of the momentum distributions (a-f) shows different dynamics of the system, solely dependent on the choice of $n$. The dashed triangle (a-c) denotes the decrease of the first Brillouin zone, corresponding to more dense lattice sites. The parameters used in the calculations are $\omega=0.08,V_0=-6$ and $g=10$.}
	\label{fig:matter-wave}
\end{figure}

The analysis of the infinitely many possible OLs is beyond the scope of this paper.
However, we will briefly discuss several key examples for $\mathbb{\Lambda}_6$.
In Fig.~\ref{fig:matter-wave}, the matter-wave interference patterns are plotted that correspond to the momentum distribution of the BEC.
First and foremost, we achieve increased lattice periodicity (closer Bragg peaks in momentum space), by choosing larger value of $n$.
The density of the lattice sites ranges from dense Fig.~\ref{fig:matter-wave}(a) ($n=1$), to intermediate Fig.~\ref{fig:matter-wave}(b) ($n=7$), to sparse Fig.~\ref{fig:matter-wave}(c) ($n=13$).
Second, contrasting to the regular triangular OLs, Fig.~\ref{fig:matter-wave}(d) shows additional interference peaks inside the first Brillouin zone for $n=67$, revealing auxiliary dynamics of the system.
These stem from emerging secondary lattice sites of the OL.
Moreover, more exotic OLs which result in a moir\'e superlattice with lattice mismatch are shown for $n=109$ in Fig.~\ref{fig:matter-wave}(e).
Finally, in Fig.~\ref{fig:matter-wave}(f), the distribution of the peaks displays a clear twelvefold rotational symmetry for $n=181$, with the distinctive structure of a quasicrystal.

Note that the OLs in Fig.~\ref{fig:matter-wave} all rely on triangular base patterns ($m=6$).
Switching symmetry by fixing a different $m$ will open rich families of OLs for exploration.
For example, setting $m=5$ in Eq.~\eqref{eq:points} will result in a wide range of quasi-periodic lattices with tenfold rotational symmetry.

\textbf{Phase synchronization} --
The main advantage of relying only on wavevector orientations to determine the OL symmetry and periodicity is that switching between different lattices amounts to activating the desired laser beams ($\mathcal{E}_j=1$ in Eq.~\eqref{eq:potential_classic}) and deactivating others.
This has the benefit of not relying on changing individual beam orientations or the operating wavelength to tune the lattice periodicity.
Similarly, spatially moving the OL can be achieved without reorienting the wavevector components, but by applying structured phase shifts $\varphi_j$ (see Eq.~\eqref{eq:potential_new}) to the input beams.
\begin{figure}[!ht]
	\centering
	\includegraphics[width=\linewidth]{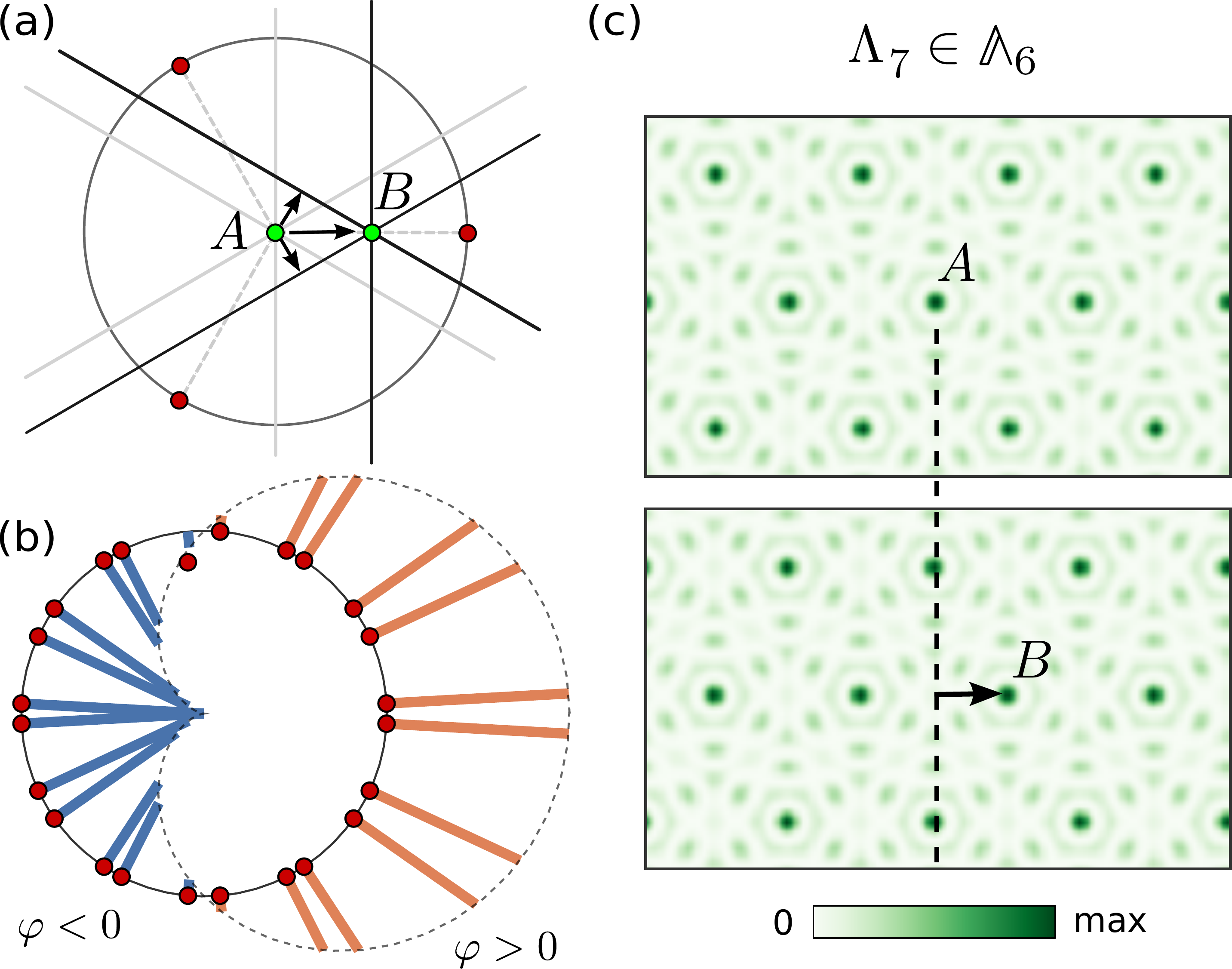}
	\caption{Phase tuning scheme for spatial translation of the OL. A diagram of the phase shifts (a) for refocusing of the constructive interference of a system with three input beams (red dots) from point $A$ to point $B$. The phase shift can be thought of as a displacement of the plane wavefront; initial (gray lines) and displaced (black lines). Each phase shift has magnitude $\varphi_j$. Positive (orange bars) or negative (blue bars) phase shifts (b), when synchronized to follow a cardioid curve (dashed line), enable the spatial translation of the OL (c).}
	\label{fig:cardioid}
\end{figure}

Displacing the superpattern by shifting each wave component along the wavevector is a known result from moir\'e theory~\cite{Amidror_2009}, and can be readily applied to the beam components.
The phase tuning scheme relies on synchronizing the displacement of the plane wave components such that the wavefronts maintain the interference pattern at each point in space.
This is illustrated in Fig.~\ref{fig:cardioid}(a).
The structured phase shifts have a magnitude that is determined by the projection of the wavevectors onto the direction of the desired displacement.
For example, horizontal motion of the moir\'e superpattern requires the following phase shifts:
\begin{equation}
	\label{eq:cardioid}
	\varphi_j = s\cdot\cos(\arg(k_j))\,.
\end{equation}
The total displacement of the OL can be tuned with the scaling factor $s$.
Visualizing $\varphi_j$ in a radial plot unveils the cardioid envelope (dashed line in Fig.~\ref{fig:cardioid}(b)).
Displacing the OL in arbitrary directions is thus achieved by orienting the cardioid curve along the corresponding axis.

Tuning the phase of individual beams is overall challenging in an experimental setting.
However, advances in the spatial light modulator technologies have already shown that tuning a large parameter space is feasible, for example in arrays of optical tweezers for cold-atom experiments~\cite{barredo_2016}.
In comparison, the integer lattice method greatly reduces the number of parameters that need to be tuned to generate and move complex OLs.
Moreover, tuning the phase can be avoided altogether for static OLs, since the prerequisite in the integer lattice method is that all laser beams are in phase, which can be readily achieved with a binary amplitude mask.

\textbf{Conclusion} --
We describe a design scheme for OLs for ultracold atom research that is predicted to give rise to rich distributions of particle momenta.
These distributions, characterized by the localized Bragg peaks, are found by numerically solving the Gross–Pitaevskii equation.

By recognizing that the OL symmetries are intimately linked to prime number distributions according to the integer lattice method, the possible ground states of the system are identified by extending the sieve of Eratosthenes algorithm to the norm of the wavevector orientations.
This approach covers the known \mbox{(quasi-)periodic} OLs and uncovers a wide range of possible OL configurations, previously unexplored to our knowledge in the scope of cold-atom physics.

The wavevector orientations of the input laser beams calculated using the integer lattice method are linked to moir\'e theory such that tuning the lattice periodicity can be achieved by simply switching the laser beams on or off, without the need to change their operating wavelength.
Therefore, the method lends itself to the dynamic study of many-body systems under varying periodicity of the carrier OL.
In addition, the OL can be continuously displaced by introducing a structured phase across the input beams, further highlighting the utility of the integer lattice method as the go-to tool for designing OLs for future cold-atom experiments.

Although finding the most suitable experimental realization is a key future challenge, the discussed beam arrangements lend themselves to be generated using standard experimental techniques.
As such, it offers an exciting framework for controlled studies of (quasi-)periodic systems of ultracold atoms, a major topic of current research.
In addition, the two-dimensional nature of the theoretical framework enables embedding the generated OLs in planar on-chip next-generation quantum simulation devices.

\begin{acknowledgments}
    We would like to thank Prof. Jacques Tempere for the helpful discussions.
    We acknowledge grant support from FWO Vlaanderen (No. 1SC0321N) to D.K.
    This work is part of a project that has received funding from the European Research Council (ERC) under the European Union's Horizon 2020 research and innovation programme (Grant Agreement No. 805222).
\end{acknowledgments}

\bibliographystyle{apsrev4-2}
\end{document}